\documentclass[aps,prd,onecolumn,preprintnumbers,10pt]{revtex4-2}
\pdfoutput=1
\usepackage{graphicx}
\usepackage{slashed}
\usepackage[dvipsnames,table]{xcolor}
\usepackage{xspace}
\usepackage[tight]{subfigure}
\usepackage{amsmath}
\usepackage{amssymb}
\usepackage{amsfonts}
\usepackage{mathrsfs}
\usepackage{comment}
\usepackage{afterpage}
\usepackage{verbatim}
\usepackage{booktabs}
\usepackage{array}
\usepackage[title,titletoc]{appendix}
\usepackage{tabularx}
\usepackage{hyperref} 

\newcolumntype{C}[1]{>{\centering\arraybackslash}p{#1}}

\def\refse#1{\mbox{Section~\ref{#1}}}
\def\refses#1{\mbox{Sections~\ref{#1}}}

\def\citere#1{\mbox{Ref.~\cite{#1}}}
\def\citeres#1{\mbox{Refs.~\cite{#1}}}

\newcommand{\newc}{\newcommand}
\newc{\beq}{\begin{equation}}
\newc{\eeq}{\end{equation}}
\newc{\bit}{\begin{itemize}}
\newc{\eit}{\end{itemize}}
\newc{\ben}{\begin{enumerate}}
\newc{\een}{\end{enumerate}}
\newc{\bce}{\begin{center}}
\newc{\ece}{\end{center}}
\newc{\bfi}{\begin{figure}}
\newc{\efi}{\end{figure}}

\newcommand{\rd}{\mathrm d}

\newcommand{\rT}{{\mathrm{T}}}

\newcommand{\rL}{{\mathrm{L}}}

\newcommand{\MeV}{\ensuremath{\,\text{MeV}}\xspace}
\newcommand{\GeV}{\ensuremath{\,\text{GeV}}\xspace}
\newcommand{\TeV}{\ensuremath{\,\text{TeV}}\xspace}

\newcommand{\PH}{\ensuremath{\text{H}}\xspace}

\newcommand{\Pp}{\ensuremath{\text{p}}}
\newcommand{\Pe}{\ensuremath{\text{e}}\xspace}

\newcommand{\Pt}{\ensuremath{\text{t}}\xspace}
\newcommand{\Pu}{\ensuremath{\text{u}}\xspace}
\newcommand{\Pd}{\ensuremath{\text{d}}\xspace}
\newcommand{\Ps}{\ensuremath{\text{s}}\xspace}
\newcommand{\Pc}{\ensuremath{\text{c}}\xspace}
\newcommand{\Pg}{\ensuremath{\text{g}}}

\newcommand{\PW}{\ensuremath{\text{W}}\xspace}
\newcommand{\PZ}{\ensuremath{\text{Z}}\xspace}

\newcommand{\Mt}{\ensuremath{m_\Pt}\xspace}
\newcommand{\MH}{\ensuremath{M_\PH}\xspace}
\newcommand{\MWOS}{\ensuremath{M_\PW^\text{OS}}\xspace}
\newcommand{\MW}{\ensuremath{M_\PW}\xspace}
\newcommand{\MZOS}{\ensuremath{M_\PZ^\text{OS}}\xspace}
\newcommand{\MZ}{\ensuremath{M_\PZ}\xspace}

\newcommand{\Gt}{\ensuremath{\Gamma_\Pt}\xspace}
\newcommand{\GH}{\ensuremath{\Gamma_\PH}\xspace}

\newcommand{\GZOS}{\ensuremath{\Gamma_\PZ^\text{OS}}\xspace}

\newcommand{\GWOS}{\ensuremath{\Gamma_\PW^\text{OS}}\xspace}

\newcommand{\GF}{\ensuremath{G_\mu}}
\newcommand{\alphas}{\ensuremath{\alpha_\text{s}}\xspace}

\newcommand{\recola}{{\sc Recola}\xspace}

\newcommand{\OpenLoops}{O\protect\scalebox{0.8}{PEN}L\protect\scalebox{0.8}{OOPS}\xspace}

\newcommand{\collier}{{\sc Collier}\xspace}

\newcolumntype{.}{D{.}{.}{-1}}
\newcolumntype{d}[1]{D{.}{.}{#1}}
\colorlet{tableoverheadcolor}{gray!37.5}
\colorlet{tableheadcolor}{gray!25}
\colorlet{tablerowcolor}{gray!12.5}

\def\draftdate{\relax}
\def\mda{\relax}
\def\mua{\relax}
\def\mla{\relax}
\def\draft{
\def\thtystars{******************************}
\def\sixtystars{\thtystars\thtystars}
\typeout{}
\typeout{\sixtystars**}
\typeout{* Draft mode!
         For final version remove \protect\draft\space in source file *}
\typeout{\sixtystars**}
\typeout{}
\def\draftdate{\today}
\def\mua{\marginpar[\boldmath\hfil$\uparrow$]%
                   {\boldmath$\uparrow$\hfil}\color{black}%
                    \typeout{marginpar: $\uparrow$}\ignorespaces}
\def\mda{\color{red}\marginpar[\boldmath\hfil$\downarrow$]%
                   {\boldmath$\downarrow$\hfil}%
                    \typeout{marginpar: $\downarrow$}\ignorespaces}
\def\mla{\marginpar[\boldmath\hfil$\rightarrow$]%
                   {\boldmath$\leftarrow $\hfil}%
                    \typeout{marginpar: $\leftrightarrow$}\ignorespaces}
\def\Mua{\marginpar[\boldmath\hfil$\Uparrow$]%
                   {\boldmath$\Uparrow$\hfil}\color{black}%
                    \typeout{marginpar: $\uparrow$}\ignorespaces}
\def\Mda{\color{red}\marginpar[\boldmath\hfil$\Downarrow$]%
                   {\boldmath$\Downarrow$\hfil}%
                    \typeout{marginpar: $\downarrow$}\ignorespaces}
\def\Mla{\marginpar[\boldmath\hfil\textcolor{red}{$\Rightarrow$}]%
                   {\boldmath\textcolor{red}{$\Leftarrow $}\hfil}%
                    \typeout{marginpar: $\leftrightarrow$}\ignorespaces}
\overfullrule 5pt
\oddsidemargin 15mm
\marginparwidth 29mm
}

\newcommand{\mc}{\mathcal}

\newcommand{\pt}[1]{p_{\rT,{#1}}}

\newcommand{\nnb}{\nonumber}

\newcommand{\GP}[1]{{{\color{blue}{[#1]}}}}

\newcommand{\moca}{{\scshape{MoCaNLO}}}
\newcommand{\stri}{{\scshape{STRIPPER}}}

\begin{document}

\title{Precise predictions for joint polarisation fractions in WZ production at the LHC} 
\author{Giovanni Pelliccioli}\email{giovanni.pelliccioli@unimib.it}   
\affiliation{Universit\`a degli Studi di Milano--Bicocca, Dipartimento di Fisica and\\ INFN Sezione di Milano--Bicocca, Piazza della Scienza 3, 20126 Milano, Italy}  
\author{Rene Poncelet}\email{rene.poncelet@ifj.edu.pl}
\affiliation{Institute of Nuclear Physics, ul. Radzikowskiego 152, 31--342 Krakow, Poland}

\begin{abstract}
We achieve for the first time NNLO QCD + NLO EW accuracy for doubly polarised $\PW\PZ$ inclusive production at the LHC, in the case of fully leptonic decays. 
Additionally, we provide estimates for missing higher-order uncertainties in QCD associated with doubly polarised differential cross sections and joint polarisation fractions, obtained both with standard scale variations and with a theory-nuisance-parameter approach.
The study is carried out in the fiducial setup of a recent ATLAS analysis of Run-2 data. 
\end{abstract}
\keywords{LHC, polarisation, NNLO QCD, NLO EW}
\preprint{COMETA-2025-50, IFJPAN-IV-2025-22}

\vspace*{0.5cm}
\maketitle


\section{Introduction}\label{sec:intro}

The study of di-boson production is a key aim of the Large Hadron Collider (LHC) analysis program, particularly with the full Run-2 and Run-3 datasets.
Such a process offers direct insights into the non-abelian nature of the electroweak (EW) interactions in the Standard Model (SM), and is therefore sensitive to potential new-physics effects on the triple-gauge couplings (TGC).
The polarisation structure of EW-boson pairs at TeV-scale energies provides a strong test of the electroweak-symmetry-breaking (EWSB) mechanism, owing to its direct connection to the longitudinal polarisation mode of $\PW$ and $\PZ$ bosons.

Amongst di-boson channels at the LHC, $\PW\PZ$ production and subsequent leptonic decays (three charged leptons and missing energy) are very well suited for polarisation studies.
The sizable production rate and the possibility to reconstruct the final state nearly completely (single neutrino) make it a good candidate for both inclusive and differential measurements.
The polarisation structure of $\PW\PZ$ inclusive production in hadron collisions has been investigated with Run-2 LHC data in the so-called \emph{polarised-template} approach \cite{Aaboud:2019gxl, CMS:2021icx, ATLAS:2022oge, ATLAS:2024qbd}.
Such analyses rely on Standard-Model (SM) predictions for intermediate EW bosons with fixed helicity states, which have become recently available thanks to a remarkable effort from the LHC theory community \cite{Ballestrero:2017bxn, BuarqueFranzosi:2019boy, Ballestrero:2019qoy, Ballestrero:2020qgv, Denner:2020bcz, Denner:2020eck, Poncelet:2021jmj, Denner:2021csi, Le:2022lrp, Le:2022ppa, Pellen:2021vpi, Hoppe:2023uux, Pelliccioli:2023zpd, Denner:2023ehn, Dao:2023kwc, Javurkova:2024bwa, Denner:2024tlu, Dao:2024ffg, Carrivale:2025mjy, Haisch:2025jqr}.
Specifically for $\PW\PZ$ production in the fully leptonic decay channel, the state-of-the-art predictions in the SM reach next-to-leading order (NLO) QCD \cite{Denner:2020eck} and NLO EW accuracy \cite{Le:2022lrp, Le:2022ppa}, including parton-shower (PS) effects via matching and merging at approximate \cite{Hoppe:2023uux} and exact \cite{Pelliccioli:2023zpd} NLO QCD accuracy. 
The polarisation structure of $\PW\PZ$ has also been studied for boosted topologies in the semi-leptonic decay channel \cite{Denner:2022riz}.
Recently, the first predictions in the SMEFT framework have been carried out at NLOPS \cite{Haisch:2025jqr} for anomalous triple-gauge couplings.

It has been shown for $\PZ\PZ$ inclusive production that higher-order QCD corrections beyond NLO are mandatory to properly model the high-energy tails of certain differential distributions for polarised signals \cite{Carrivale:2025mjy}, either via multi-jet merging or through exact next-to-next-to-leading-order (NNLO) QCD corrections.
Owing to the strong interest that $\PW\PZ$ has received from an experimental viewpoint in the context of polarisation measurements \cite{Aaboud:2019gxl, CMS:2021icx, ATLAS:2022oge, ATLAS:2024qbd}, it is of paramount importance to include NNLO QCD effects, possibly combined with NLO EW ones, in the polarisation modelling of $\PW\PZ$ inclusive production. 
This is the first achievement of this work, bringing the SM description of doubly polarised states to the same level of perturbative accuracy as the full off-shell process \cite{Grazzini:2016swo, Grazzini:2017ckn, Biedermann:2017oae, Kallweit:2019zez, Chiesa:2020ttl, Lindert:2022qdd}.

Additionally, it is crucial to provide sound estimates for missing higher-order uncertainties (MHOU) in polarised predictions, particularly for polarisation fractions, in light of comparisons with real experimental data.
It became customary to use the variation of unphysical scales that arise in perturbative computations to estimate the size of the higher-order terms.
Considering polarisation fractions, which are defined as ratios of cross sections, it is crucial to consider the correlations between numerator and denominator.
Scale variations are known to show pathological behaviour in such cases, which can lead to an underestimation of the theory uncertainty.
Theory nuisance parameters \cite{Tackmann:2024kci} provide an alternative way to estimate MHOU and have been recently investigated in various scenarios \cite{McGowan:2022nag, Lim:2024nsk, Cridge:2025wwo, Chang:2025ohh}.
In this work, we employ the approach outlined in \cite{Lim:2024nsk} to derive MHOU estimates and compare them with those from the conventional scale-variation approach.

The structure of the paper is as follows. 
After depicting the details of the calculation in~\refse{sec:calc}, we show integrated and differential results respectively in \refses{sec:Iresults} and~\ref{sec:Dresults}, considering fiducial ATLAS selections. A discussion of joint polarisation fractions is carried out in \refse{sec:Fresults}. The MHOU estimates are presented for differential polarised results both in \refse{sec:Dresults} and \refse{sec:Fresults}. In \refse{sec:conclusion} we draw our conclusions.

\section{Details of the calculation}\label{sec:calc}
We consider $\PW^{+}\PZ$ inclusive production at the LHC in the three-lepton channel:
\beq\label{eq:processdef}
  \Pp\Pp\rightarrow \PW^+\,(\rightarrow \Pe^+\nu_{\Pe})\,\PZ\,(\rightarrow{\mu^+\mu^-})
  + X\,.
\eeq
The doubly polarised signals are computed in the double-pole approximation (DPA)\cite{Stuart:1991cc, Stuart:1991xk, Aeppli:1993cb, Aeppli:1993rs, Denner:2005fg, Denner:2019vbn} and by selecting individual polarisation states in all tree-level, one-loop, and two-loop SM amplitudes entering the calculation.
This represents the most common strategy used for polarised-boson computations \cite{Ballestrero:2017bxn, BuarqueFranzosi:2019boy, Ballestrero:2019qoy, Ballestrero:2020qgv, Denner:2020bcz, Denner:2020eck, Poncelet:2021jmj, Denner:2021csi, Le:2022lrp, Le:2022ppa, Pellen:2021vpi, Hoppe:2023uux, Pelliccioli:2023zpd, Denner:2023ehn, Dao:2023kwc, Javurkova:2024bwa, Denner:2024tlu, Dao:2024ffg, Carrivale:2025mjy, Haisch:2025jqr}. 
The polarisation states are defined in the di-boson centre-of-mass (CM) frame, following the same choice as in ATLAS measurements \cite{Aaboud:2019gxl, ATLAS:2022oge, ATLAS:2024qbd} and in previous theory calculations \cite{Denner:2020eck, Le:2022lrp, Le:2022ppa}.

The calculation relies on two Monte Carlo (MC) codes, \moca~and \stri.

\moca~is a general-purpose MC integration code that has been used to calculate NLO QCD + EW predictions for polarised-boson-pair inclusive production \cite{Denner:2020bcz, Denner:2020eck, Denner:2021csi, Denner:2022riz, Denner:2023ehn, Carrivale:2025mjy} and scattering \cite{Denner:2024tlu}.
It is interfaced with the \recola1 SM-amplitude library \cite{Actis:2012qn, Actis:2016mpe} and the \collier~library for one-loop reduction \cite{Denner:2016kdg}.
The QCD and QED infrared (IR) singularities at NLO are handled in the dipole formalism \cite{Catani:1996vz, Dittmaier:1999mb, Catani:2002hc, Basso:2015gca}.
Polarised-boson calculations in \moca~rely on a general implementation of the pole approximation \cite{Stuart:1991cc, Stuart:1991xk, Aeppli:1993cb, Aeppli:1993rs, Denner:2005fg, Denner:2019vbn} and on the separation of helicity contributions at the level of EW-boson propagators in the amplitudes (see \citeres{Denner:2021csi, Denner:2024tlu} for technical details). 
 
\stri~is a C++ implementation of the four-dimensional sector-improved residue subtraction scheme \cite{Czakon:2010td, Czakon:2014oma, Czakon:2019tmo} which automates the subtraction of IR QCD singularities and performs the numerical MC integration of differential cross sections at NNLO in QCD.
The implementation supports intermediate polarisations of EW bosons in the pole and narrow-width approximations, and has been applied to several polarisation studies \cite{Poncelet:2021jmj, Pellen:2021vpi, Pellen:2022fom} with $\PW^+\PW^-$, $\PZ\PZ$ and $\PW+\text{jet}$ final states.
Tree-level amplitudes for Born, single-, and double-real-radiation contributions are provided by the \textsc{AvH} library \cite{Bury:2015dla}, while the one-loop amplitudes are obtained from \OpenLoops2 \cite{Cascioli:2011va, Buccioni:2017yxi, Buccioni:2019sur}.
The two-loop amplitudes for boson-pair production rely on the \textsc{VVamp} library \cite{Gehrmann:2015ora}.

In both codes, the intermediate (un)polarised bosons are treated in the DPA following the conventions of \citere{Denner:2021csi} for both on-shell-projection mappings and polarisation-vector definitions.
For further details on the DPA, we refer to \citeres{Denner:2021csi, Denner:2024tlu}.

We have performed several validation checks.  In particular, a complete comparison between \moca~and \stri~has been carried out at NLO QCD accuracy for both integrated and differential results, finding perfect agreement.
A more in-depth comparison of these MC tools with other generators has been recently carried out for $\PZ\PZ$ inclusive production \cite{Carrivale:2025mjy}.
The polarisation fractions at NLO QCD + EW (NLO$^{(+)}$) from our calculation are found to be in good agreement with those of \citere{Le:2022lrp}.
The numerical results of this comparison are shown in Appendix~\ref{app:compNinh}.
Additionally, the polarised one-loop amplitudes from a \emph{ad-hoc} version of \OpenLoops2 have been validated against \recola1 at the phase-space-point level.

The calculation is performed in the five-flavour scheme, and a unit CKM matrix is assumed.
All quarks (but the top) and leptons have vanishing mass. 
The pole masses and widths of weak bosons are calculated by converting \cite{Bardin:1988xt} the corresponding on-shell PDG values \cite{ParticleDataGroup:2022pth},
\begin{alignat}{2}\label{eq:ewmasses}
 \MWOS &= 80.377  \GeV,&\qquad 
 \GWOS &= 2.085\GeV,\nnb\\
 \MZOS &= 91.1876  \GeV,& \qquad
 \GZOS &= 2.4952 \GeV.
\end{alignat}
The $G_\mu$ scheme \cite{Denner:2000bj} is used with the Fermi constant set to  $\GF = 1.16638\cdot 10^{-5} \GeV^{-2}$\,. The EW coupling is extracted via
\beq
\alpha = \frac{\sqrt{2}}{\pi}\,G_\mu\MW^2\left(1-\frac{\MW^2}{\MZ^2}\right)\,.
\eeq
The top-quark and Higgs-boson masses and widths read,
\begin{alignat}{2}\label{eq:thmasses}
 \Mt &= 172.69\GeV,&\qquad \Gt &= 1.36\GeV, \nnb\\
 \MH &= 125.25\GeV,&\qquad \GH &= 4.07\MeV. 
\end{alignat}
We use \sloppy{\tt NNPDF31\_nnlo\_as\_0118\_luxqed} \cite{Ball:2017nwa, Bertone:2017bme} parton-distribution functions (PDFs), provided to the Monte Carlo codes via the LHAPDF interface \cite{Buckley:2014ana}.
Also, the running of the strong coupling constant $\alphas$ is evaluated with built-in LHAPDF routines.
The $\overline{\rm MS}$ factorisation scheme is employed for the treatment of initial-state collinear singularities.
The central factorisation and renormalisation scales are set to the fixed value,
\beq\label{eq:muF_muR}
 \mu_{\rm F}=\mu_{\rm R}=\frac{\MW+\MZ}2\,,
\eeq
following the same choice as previous studies \cite{Denner:2020eck,Le:2022lrp}.
The MHOU from scale variations are extracted from seven-point variations of $\mu_{\rm F}$ and $\mu_{\rm R}$.

Following the fiducial ATLAS event selections \cite{ATLAS:2022oge,ATLAS:2024qbd} we apply the following cuts:
\begin{eqnarray}\label{eq:fid}
&\pt{\Pe^+(\mu^\pm)}>20\,(15)\GeV,\qquad
M_{\rT,\PW}>30\GeV,\qquad 
81\GeV<M_{\mu^+\mu^-}<101\GeV,&\nnb\\
&
|\eta_{\Pe^+(\mu^\pm)}|<2.5,\qquad\Delta R_{\Pe^+\mu^\pm}>0.3,\qquad \Delta R_{\mu^+\mu^-}>0.2,&
\end{eqnarray}
where $M_{\rT,\PW} = \sqrt{2\,\pt{\Pe^+}\pt{\rm mis}(1-\cos\Delta\phi_{\Pe^+,\rm mis})}\,$.
The QCD partons are clustered into jets through the anti-$k_{\rm t}$ algorithm \cite{Cacciari:2008gp} with resolution radius $R=0.4$, while photons are clustered with charged particles with cone dressing and resolution radius $R=0.1$.

\section{Results}
\subsection{Fiducial cross sections}\label{sec:Iresults}
In this section, we show fiducial results at fixed order for the ATLAS setup \cite{ATLAS:2022oge}. 
For the first time, we set the new state-of-the-art perturbative accuracy of SM predictions for $\PW\PZ$ production and decay in the pole approximation and for intermediate polarised EW bosons, matching the same accuracy of the state-of-the-art for the full off-shell modelling \cite{Kallweit:2019zez}. 
We provide predictions at NLO and NNLO accuracy. At NLO, we have QCD and EW corrections, which we decompose as:
\begin{eqnarray*}
\rd\sigma_{\rm NLO}^{\rm QCD} &=& \rd\sigma_{{\rm LO}}\,\left( 1+\delta^{\rm (1)}_{{\rm QCD},\,\Pu\bar{\Pd}}\right)\,
+ \rd\sigma_{q\Pg}  \,=\,
\rd\sigma_{{\rm LO}}\,\left(1+\delta^{\rm (1)}_{{\rm QCD}}\right),\nnb\\
\rd\sigma_{\rm NLO}^{\rm EW} &=& \rd\sigma_{{\rm LO}}\,\left( 1+\delta^{\rm (1)}_{{\rm EW},\,\Pu\bar{\Pd}}\right)\,
+ \rd\sigma_{q\gamma}  \,=\,
\rd\sigma_{{\rm LO}}\,\left(1+\delta^{\rm (1)}_{{\rm EW}}\right)\,.\label{eq:defNLO}
\end{eqnarray*}
Similarly, we write the NNLO QCD corrections as
\begin{eqnarray}
\rd\sigma_{\rm NNLO}^{\rm QCD} &=& \rd\sigma_{{\rm LO}}\,\left( 1+\delta^{\rm (1)}_{{\rm QCD},\,\Pu\bar{\Pd}}+\delta^{\rm (2)}_{{\rm QCD},\,\Pu\bar{\Pd}}\right)   
+ \rd\sigma_{q\Pg}\,(1+\delta^{\rm (1)}_{{\rm QCD},\,q\Pg}) 
+ \rd\sigma_{qq'}
+ \rd\sigma_{\Pg\Pg}
\,=\,
\rd\sigma_{{\rm LO}}\,\left(1+\delta^{\rm (1)}_{{\rm QCD}}+\delta^{\rm (2)}_{{\rm QCD}}\right)\,.
\label{eq:defNNLO}
\end{eqnarray}
We then combine (N)NLO QCD and NLO EW corrections with both an additive and a multiplicative approach, according to:
\begin{eqnarray}
\rd\sigma_{\rm NLO}^{(+)} &=& \rd\sigma_{{\rm LO}}\,\left( 1+\delta^{\rm (1)}_{{\rm QCD},\,\Pu\bar{\Pd}}+\delta^{\rm (1)}_{{\rm EW},\,\Pu\bar{\Pd}}\right)\,
+ \rd\sigma_{q\Pg} + \rd\sigma_{q\gamma}  , 
\nnb\\  
\rd\sigma_{\rm NLO}^{(\times)} &=& \rd\sigma_{{\rm LO}}\,\left( 1+\delta^{\rm (1)}_{{\rm QCD},\,\Pu\bar{\Pd}}\right)\,\left(1+\delta^{\rm (1)}_{{\rm EW},\,\Pu\bar{\Pd}}\right) 
+ \rd\sigma_{q\Pg} + \rd\sigma_{q\gamma} ,  
\nnb\\
\rd\sigma_{\rm NNLO}^{(+)} &=& \rd\sigma_{{\rm LO}}\,\left( 1+\delta^{\rm (1)}_{{\rm QCD},\,\Pu\bar{\Pd}}+\delta^{\rm (2)}_{{\rm QCD},\,\Pu\bar{\Pd}}+\delta^{\rm (1)}_{{\rm EW},\,\Pu\bar{\Pd}}\right)   
+ \rd\sigma_{q\Pg}\,(1+\delta^{\rm (1)}_{{\rm QCD},\,q\Pg}) 
+ \rd\sigma_{qq'}
+ \rd\sigma_{\Pg\Pg}
+ \rd\sigma_{q\gamma}   ,
\nnb\\  
\rd\sigma_{\rm NNLO}^{(\times)} &=& \rd\sigma_{{\rm LO}}\,\left( 1+\delta^{\rm (1)}_{{\rm QCD},\,\Pu\bar{\Pd}}+\delta^{\rm (2)}_{{\rm QCD},\,\Pu\bar{\Pd}}\right)\,\left(1+\delta^{\rm (1)}_{{\rm EW},\,\Pu\bar{\Pd}}\right)     
+ \rd\sigma_{q\Pg}\,(1+\delta^{\rm (1)}_{{\rm QCD},\,q\Pg}) 
+ \rd\sigma_{qq'}
+ \rd\sigma_{\Pg\Pg} 
+ \rd\sigma_{q\gamma}\,.\label{eq:mult_add} 
\end{eqnarray}

\begin{table*}
  \begin{center}
    \begin{tabular}{ccccccccc}
      \hline\rule{0ex}{2.7ex}
      \cellcolor{yellow!9} state  
      & \cellcolor{yellow!9} $\sigma_{\rm LO}$ (fb) 
      & \cellcolor{yellow!9} $\delta^{\rm (1)}_{\rm QCD}$
      & \cellcolor{yellow!9} $\delta^{\rm (2)}_{\rm QCD}$
      & \cellcolor{yellow!9} $\delta^{\rm (1)}_{\rm EW}$ 
      & \cellcolor{yellow!9} $\sigma^{(+)}_{\rm NLO}$ (fb)
      & \cellcolor{yellow!9} $\sigma^{(\times)}_{\rm NLO}$ (fb)
      & \cellcolor{yellow!9} $\sigma^{(+)}_{\rm NNLO}$ (fb)
      & \cellcolor{yellow!9} $\sigma^{(\times)}_{\rm NNLO}$ (fb)\\[0.1cm]
      \hline\\[-0.15cm]
off-sh. & $ 19.869  (2  )^{+4.6  \%} _{-5.7  \%}  $ &   $ +78.6   \% $ &  $ +16.4  \% $ &  $ -4.4   \% $ &   $ 34.60  (1  )  ^{+  5.4  \%} _{ -4.3  \%} $ &  $ 34.19  (1 )  ^{+  5.3  \%} _{ -4.2  \%} $ &  $ 37.87(13)  ^{+  2.5  \%} _{ -2.2  \%} $ & $ 37.50(13)  ^{+ 2.7  \%} _{-2.2  \%} $ \\[0.25cm] 
unp. & $ 19.457  (1  )^{+4.6  \%} _{-5.7  \%}  $ &   $ +79.0   \% $ &  $ +16.3  \% $ &  $ -4.4   \% $ &   $ 33.957  (6  )  ^{+  5.4  \%} _{ -4.3  \%} $ &  $ 33.550  (6  )  ^{+  5.3  \%} _{ -4.3  \%} $ &  $ 37.13  (8  ) ^{+  2.6  \%} _{ -2.2  \%} $ & $ 36.77(8)  ^{+ 2.7  \%} _{-2.3  \%} $ \\[0.25cm] 
LL & $ 1.5326  (1  )^{+5.0  \%} _{-6.1  \%}  $ &   $ +29.8   \% $ &  $ +11.1  \% $ &  $ -4.5   \% $ &   $ 1.921  (0  )  ^{+  2.8  \%} _{ -2.3  \%} $ &  $ 1.895  (0  )  ^{+  2.7  \%} _{ -2.2  \%} $ &  $ 2.091  (3  ) ^{+  2.1  \%} _{ -1.8  \%} $ &  $ 2.064(3)  ^{+ 2.4  \%} _{-1.7  \%} $ \\[0.25cm] 
LT & $ 2.0716  (2  )^{+5.6  \%} _{-6.8  \%}  $ &   $ +159.2   \% $ &  $ +42.7  \% $ &  $ -3.5   \% $ &   $ 5.295  (1  )  ^{+  7.4  \%} _{ -5.9  \%} $ &  $ 5.244  (1  )  ^{+  7.3  \%} _{ -5.9  \%} $ &  $ 6.181  (23  ) ^{+  4.2  \%} _{ -3.4  \%} $ & $ 6.134(23)  ^{+ 4.3  \%} _{-3.5  \%} $ \\[0.25cm] 
TL & $ 1.9531  (4  )^{+5.6  \%} _{-6.8  \%}  $ &   $ +161.8   \% $ &  $ +43.7  \% $ &  $ -0.9   \% $ &   $ 5.097  (1  )  ^{+  7.4  \%} _{ -5.9  \%} $ &  $ 5.046  (1  )  ^{+  7.3  \%} _{ -5.9  \%} $ &  $ 5.951  (13  ) ^{+  3.9  \%} _{ -3.3  \%} $ & $ 5.940(13)  ^{+ 3.9  \%} _{-3.3  \%} $ \\[0.25cm] 
TT & $ 13.751  (1  )^{+4.3  \%} _{-5.4  \%}  $ &   $ +60.9   \% $ &  $ +9.0  \% $ &  $ -5.1   \% $ &   $ 21.431  (5  )  ^{+  4.6  \%} _{ -3.7  \%} $ &  $ 21.167  (5  )  ^{+  4.5  \%} _{ -3.7  \%} $ &  $ 22.67  (8  ) ^{+  1.6  \%} _{ -1.5  \%} $ & $ 22.43(8)  ^{+ 1.8  \%} _{-1.6  \%} $ \\[0.25cm] 
    \hline
     \end{tabular}\qquad
  \end{center}
  \caption{ Fiducial cross sections (in fb) in the ATLAS setup \cite{ATLAS:2022oge} described in Eq.~\ref{eq:fid}. The percentages $\delta$'s are  NLO (1) and NNLO corrections (2), relative to the LO. The cross sections with $+(\times)$ labels understand an additive (multiplicative) combination of
  QCD and EW corrections, according to Eq.~\ref{eq:mult_add}.
  For doubly polarised states, the first (second) label is associated with the $\PW (\PZ)$ boson. Numbers in parentheses represent MC-integration numerical uncertainties.
  \label{tab:predIntegNNLO}
  }
\end{table*}

\begin{table*}
  \begin{center}
    \begin{tabular}{ccccccccc}
      \hline\rule{0ex}{2.7ex}
        \cellcolor{yellow!9} state
      & \cellcolor{yellow!9} $\phantom{x}\delta^{\rm (1)}_{\rm QCD,\Pu\bar{\Pd}}\phantom{x}$  
      & \cellcolor{yellow!9} $\rd\sigma_{q\Pg}/\rd \sigma_{\rm LO}$ 
      & \cellcolor{yellow!9} $\phantom{x}\delta^{\rm (2)}_{\rm QCD,\Pu\bar{\Pd}}\phantom{x}$  
      & \cellcolor{yellow!9} $\rd\sigma_{q\Pg}\delta^{\rm (1)}_{{\rm QCD},\,q\Pg}/\rd\sigma_{\rm LO}$ 
      & \cellcolor{yellow!9} $\rd\sigma_{qq'}/\rd \sigma_{\rm LO}$ 
      & \cellcolor{yellow!9} $\rd\sigma_{\Pg\Pg}/\rd \sigma_{\rm LO}$ 
      & \cellcolor{yellow!9} $\rd\sigma_{q\gamma}/\rd \sigma_{\rm LO}$ 
      & \cellcolor{yellow!9} $\phantom{x}\delta^{\rm (1)}_{\rm EW,\Pu\bar{\Pd}}\phantom{x}$\\[0.2cm]
      \hline\\[-0.15cm]
      \, LL & $ 31.9 \% $ &$ -2.1\% $&$ 7.7\% $&$ 0.2\% $&$2.7 \% $&$ 0.5\% $&   $ 0.6   \% $ &   $ -5.1   \% $ \\[0.2cm]
      \, LT & $ 50.5 \% $ &$ 108.5\% $&$ 13.3\% $&$ 19.5\% $&$10.6 \% $&$ -0.8\% $&   $ 1.4   \% $ &   $ -4.9   \% $ \\[0.2cm]    
      \, TL  & $ 50.9 \% $  & $ 110.9\% $  & $ 13.7\% $& $ 20.1\% $& $ 10.7\% $& $ -0.8\% $&   $ 4.1   \% $ &   $ -5.0   \% $ \\[0.2cm]    
      \, TT  &  $29.7 \% $& $ 31.3\% $& $ 5.7\% $& $ -0.8\% $& $ 3.9\% $& $-0.1 \% $&   $ 1.4   \% $ &   $ -6.5   \% $ \\[0.2cm]
 \hline
    \end{tabular}\qquad
  \end{center}
  \caption{QCD corrections relative to the LO in the ATLAS setup \cite{ATLAS:2022oge} described in Eq.~\ref{eq:fid}. Following Eqs.~\ref{eq:defNLO}--\ref{eq:mult_add}, the corrections are defined such that 
  $\delta^{\rm (1)}_{\rm QCD}=\delta^{\rm (1)}_{\rm QCD,\Pu\bar{\Pd}}+(\rd\sigma_{q\Pg}/\rd \sigma_{\rm LO})$,
  \sloppy
  $\delta^{\rm (2)}_{\rm QCD}=\delta^{\rm (2)}_{\rm QCD,\Pu\bar{\Pd}}+(\rd\sigma_{q\Pg}\delta^{\rm (1)}_{\rm QCD,q\Pg}/\rd \sigma_{\rm LO})+(\rd\sigma_{qq'}/\rd \sigma_{\rm LO})+(\rd\sigma_{\Pg\Pg}/\rd \sigma_{\rm LO})$,
  and
  $\delta^{\rm (1)}_{\rm EW}=\delta^{\rm (1)}_{\rm EW,\Pu\bar{\Pd}}+(\rd\sigma_{q\gamma}/\rd \sigma_{\rm LO})$.
  The first (second) polarisation label refers to the $\PW (\PZ)$ boson. \label{tab:relNNLO}
  }
\end{table*}

The $\PW\PZ$ process is characterised by large QCD corrections due to an approximate amplitude zero in the dominant $\rT\rT$ amplitudes ($\pm\mp$), which is present at LO but not any more in the presence of additional QCD radiation \cite{Ohnemus:1991gb, Baur:1994ia}. 
Furthermore, in the high-energy limit, the Goldstone equivalence theorem \cite{Cornwall:1974km, Vayonakis:1976vz, Chanowitz:1985hj, Gounaris:1986cr} implies a LO suppression of the mixed states compared to the $\rL\rL$ and $\rT\rT$ ones.
This is one source of the dramatically large corrections which appear at NLO QCD for the mixed states.
A second source of this effect is the opening of the gluon--quark partonic channels that leads to huge Sudakov logarithms \cite{Rubin:2010xp, Kallweit:2019zez} in the limit where one boson is softer than the other one.
Such effects are minor for the LL state, which favours kinematic configurations where both bosons are rather soft \cite{Denner:2020eck}.
The impressive size of the $q\Pg$ contributions to the mixed states can also be understood in a different way.
At the LHC, single EW bosons produced in association with a jet (with the $q\Pg$ channel being the dominant one) are typically left-handed \cite{Bern:2011ie} while the longitudinal mode is suppressed (vanishes at zero $\pt{V}$), owing to the spin balance between the initial and final state.
The radiation of an additional (soft) EW boson, which is longitudinal, is favoured as a transverse one would bring a non-vanishing third component of the spin, making it more difficult to achieve the spin balance. 

As previously observed in the literature \cite{Le:2022lrp, Le:2022ppa}, a marked difference between the LT and TL contributions is found at the level of the EW corrections ($-3.5\%$ for LT, $-0.9\%$ for TL).
This can be traced back to the photon-induced contribution. While the genuine EW correction to the LO partonic channels ($\Pu\bar{\Pd},\,\Pc\bar{\Ps}$) are both at the $-5\%$ level, the TL polarisation state receives a photon-induced contribution that is 3 times larger than the one contributing to LT, almost cancelling the NLO EW correction to the quark--anti-quark channel.
The numerical results are shown in the two rightmost columns of Table~\ref{tab:relNNLO}.
Compared to other polarisation modes, the LL signal receives a relatively small photon-induced contribution  ($0.6\%$).
These results highlight how the different spin structure introduced by the photon in the initial state has non-negligible effects on combined (N)NLO polarised cross sections. 

As shown in Table~\ref{tab:predIntegNNLO}, the additive and multiplicative combinations at NLO (and NNLO) accuracy differ by 1\% and feature very similar QCD-scale uncertainties.
As a rough quantitative estimate, we expect the radiative corrections of orders $\mc O (\alpha\alphas)$ and $\mc O (\alpha\alphas^2)$ to the quark--antiquark channel to be at the 1\% level.

The NNLO QCD corrections receive contributions from genuine corrections to the partonic channels already present at NLO, as well as from newly opened partonic channels. 
At variance with $\PZ\PZ$ and $\PW^+\PW^-$ production, $\PW\PZ$ does not receive PDF-enhanced contributions from one-loop $\Pg\Pg$-initiated squared diagrams.
The $\Pg\Pg$ channel gives a sub-per-cent correction from pure real-radiation diagrams to all polarisation modes.
The $\rL\rL$ and $\rT\rT$ modes receive 5\%-level corrections to the dominant $q_{\Pu}\bar{q}_{\Pd}$ channel, and $3\%$-level real corrections from newly opened $qq'$ channels. 
These polarisation modes receive sub-per-cent shifts from the NLO corrections to the $\Pg q$ channel.
On the contrary, the mixed states receive 10-to-20\% corrections from all partonic channels but the $\Pg\Pg$ one.

We anticipate that, despite sizeable (and often very large) higher-order corrections to fiducial polarised cross sections, the impact of such radiative effects is more moderate with regard to polarisation fractions, defined as ratios between polarised and unpolarised cross sections.
A broad discussion on the fractions is postponed to \refse{sec:Fresults}.

Compared to integrated cross sections, the effects from NNLO QCD and NLO EW corrections can become even larger in more exclusive phase-space regions and when looking at differential observables. This is carried out in the next subsection.

\subsection{Differential distributions}\label{sec:Dresults}
In this section, we consider differential results in two observables that show relevant features for polarisation extraction in $\PW\PZ$ events at the LHC.

\begin{figure*}
    \centering
    \subfigure[\label{fig:ptz}]{\includegraphics[width=0.48\textwidth,page=2]{plots/wz_nloqcd_for_publ.pdf}}
    \subfigure[\label{fig:dyez}]{\includegraphics[width=0.48\textwidth,page=1]{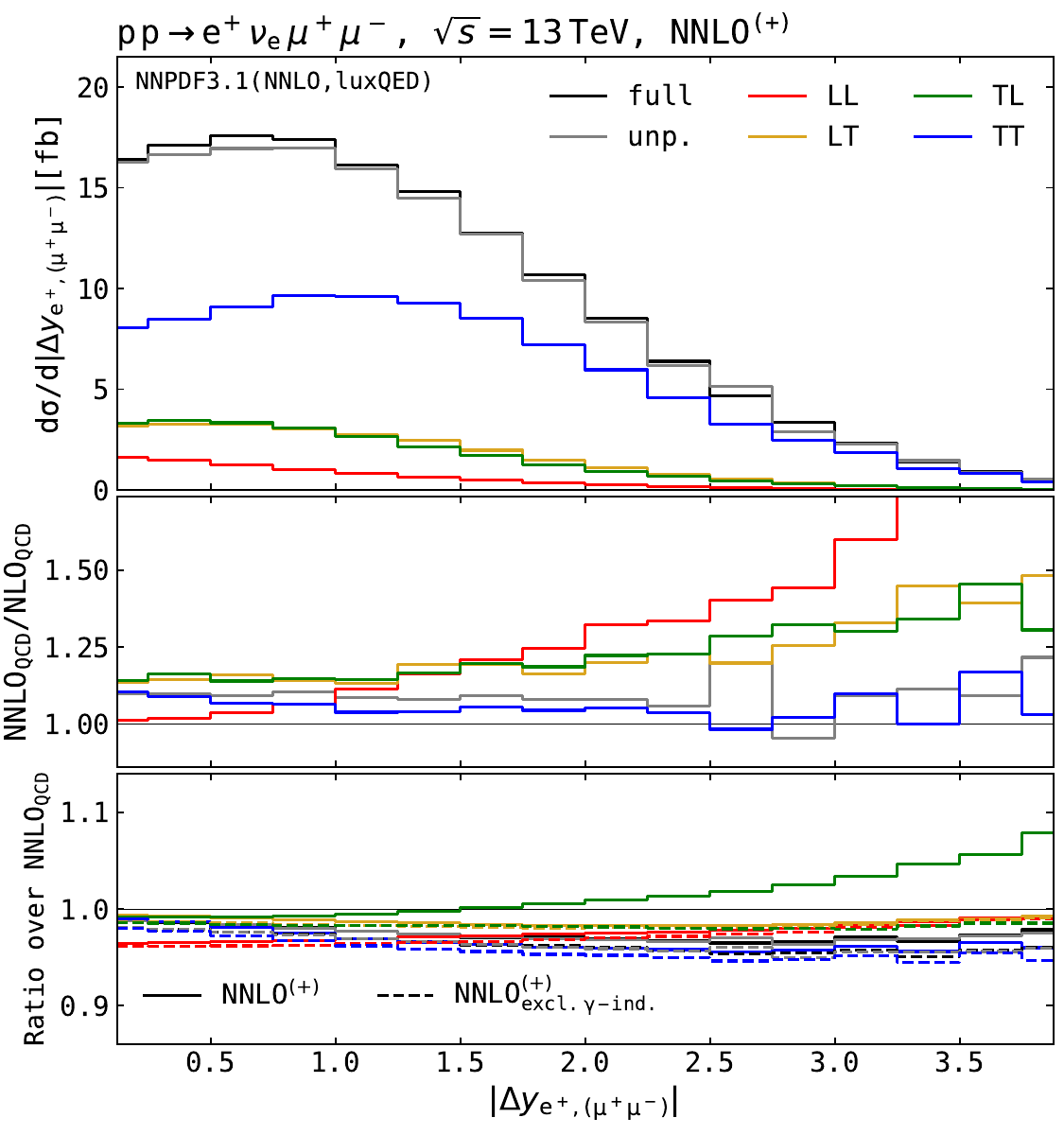}}
    \caption{Distributions in the transverse-momentum of the muon--antimuon system (a)
    and the rapidity separation between the positron and the muon--antimuon system (b) in the fiducial ATLAS setup \cite{ATLAS:2022oge}. 
    Upper panels: absolute differential cross sections at NNLO$^{(+)}$ accuracy, namely combining additively NNLO QCD and NLO EW corrections according to Eq.~\ref{eq:mult_add}. Middle panels: ratios between NNLO and NLO QCD distributions. Lower panels: ratio of NNLO$^{(+)}$ distributions including (solid) and excluding (dashed) photon-induced contributions over the NNLO QCD ones. Colour key: full off-shell (black), unpolarised (grey), $\rL\rL$ (red), $\rL\rT$ (yellow), $\rT\rL$ (green), and $\rT\rT$ (blue), where the first (second) polarisation label is associated with the $\PW\,(\PZ)$ boson.
    }
    \label{fig:diff}
\end{figure*}

The high-energy behaviour of the doubly polarised signals is probed by looking at the transverse momentum of the $\PZ$ boson, identified as the muon--antimuon system. This is considered in Fig.~\ref{fig:ptz}.
Differential measurements of this observable have been obtained by ATLAS in the most recent polarisation study of the $\PW\PZ$ channel \cite{ATLAS:2024qbd}.
In the limit of large transverse momentum, the states with a transverse $\PZ$ boson ($\rT\rT$ and $\rL\rT$) dominate over the states with a longitudinal $\PZ$ boson, which are suppressed by more than one order of magnitude at the $1\TeV$ scale. As also observed at NLO accuracy \cite{Le:2022lrp}, the $\rT\rL$ mode becomes smaller than the $\rL\rL$ one around $300\GeV$, owing to a large-$p_{\rT}$ longitudinal boson produced with another transverse boson which can have moderate-to-small $p_{\rT}$.
This is a suppressed configuration, as longitudinal bosons tend to be much softer than transverse ones in inclusive di-boson processes \cite{Denner:2020bcz, Denner:2021csi, Le:2022lrp}. 
The expected suppression of the $\rL\rT$ state at high energy is not visible as the large transverse momentum of the $\PZ$ boson is not necessarily correlated with a large invariant mass of the $\PW\PZ$ system, especially beyond LO, where hard QCD radiation creates an asymmetry in the hardness of the two bosons.
These arguments are also the origin of the larger NNLO QCD corrections found for $\rT\rL$ than for $\rL\rT$ in the asymptotic regime.
The NNLO QCD corrections do not exceed $10\%(20\%)$ in absolute value for the $\rL\rL\,(\rT\rT)$ states in the whole considered range.
Both the $\rL\rL$ and the $\rT\rT$ modes receive increasingly large and negative genuine EW corrections to the LO partonic channel.
Still, the effect is smaller for $\rT\rT$ owing to the smaller factors multiplying the leading EW Sudakov logarithms \cite{Denner:2000jv}.
The EW effects look even smaller in Fig.~\ref{fig:ptz} because they are partially washed away by the large QCD corrections for the $\rT\rT$ mode.
The NLO EW corrections to the mixed states become positive at moderate transverse momentum, in agreement with \citere{Le:2022lrp}, owing to the LO suppression.
The effect of photon-induced contributions is similarly sizeable ($\approx 5\%$ above $500\GeV$) at large transverse momentum for the three states involving transverse bosons.

An enhanced discrimination power amongst polarisation states is given by the rapidity separation between the charged lepton from the $\PW$-boson decay and the $\PZ$ boson (muon--antimuon system), which is considered in Fig.~\ref{fig:dyez}.
This quantity is an input variable both for the neural-network architecture used by ATLAS in the measurement of inclusive polarisation fractions \cite{ATLAS:2022oge} and for the boosted-decision tree employed in the more recent analysis \cite{ATLAS:2024qbd}.
The $\rL\rL$ shape has a maximum at zero and then decreases faster than other modes.
The two mixed states show very similar shapes and peak around $|\Delta y_{\Pe^+\PZ}|\approx 0.3$. The $\rT\rT$ shape peaks at $|\Delta y_{\Pe^+\PZ}|\approx 1.3$.
The combined NNLO corrections mildly change the NLO-accurate shapes, with the largest effects appearing in the suppressed region at large rapidity separation. 
The second-order QCD corrections become increasingly large towards the most suppressed regime for the $\rL\rL$ state.
The mixed states are characterised by a similar effect, although to a lesser extent.
The QCD corrections to the $\rT\rT$ distribution are flatter and tend to vanish at large rapidity separations. 
Compared to NNLO QCD distributions, the NLO EW effects are between $-1\%$ and $-6\%$ for all states.
The only exception is the $\rT\rL$ polarisation state that receives an increasingly large and positive shift from the photon-induced channels ($+8\%$ at $|\Delta y_{\Pe^+\PZ}|\approx 4$).

\begin{figure*}
    \centering
    \subfigure[\label{fig:ptzTNP}]{\includegraphics[width=0.48\textwidth,page=2]{plots/wz_TNP_kfact.pdf}}
    \subfigure[\label{fig:dyezTNP}]{\includegraphics[width=0.48\textwidth,page=1]{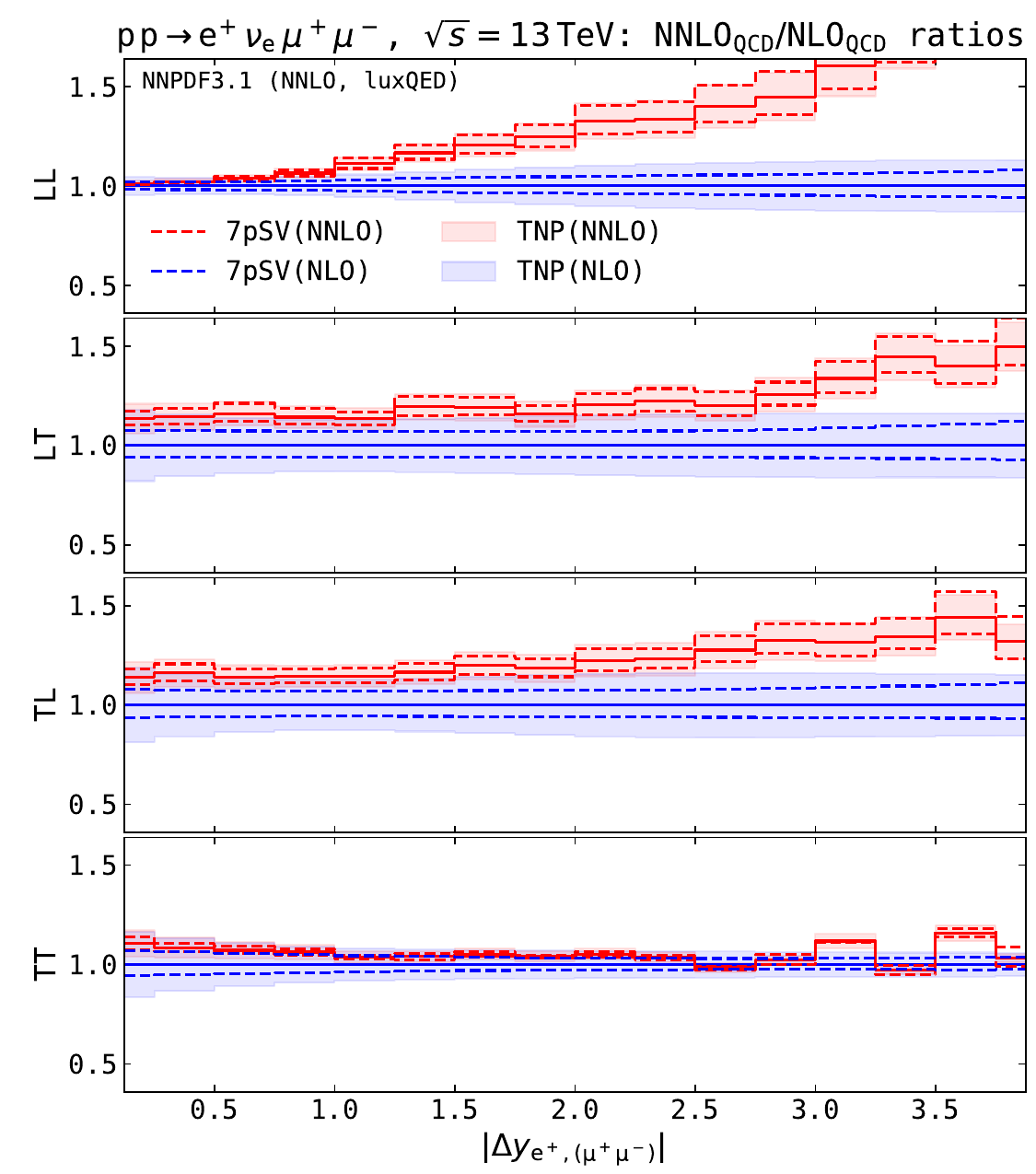}}
    \caption{
    Differential QCD $K$-factors ($\rd\sigma^{\rm QCD}_{\rm NNLO}/\rd\sigma^{\rm QCD}_{\rm NLO}$) in the transverse-momentum of the muon--antimuon system (a)
    and the rapidity separation between the positron and the muon--antimuon system (b) in the fiducial ATLAS setup \cite{ATLAS:2022oge}. 
    The NLO (NNLO) results are shown in blue (red) colour for the LL, LT, TL, and TT polarisation states, from top down. The MHOU QCD uncertainties are computed with seven-point scale variations (``7pSV'', dashed boundary lines) and with the theory nuisance parameter approach (``TNP'', shaded bands).
    }
    \label{fig:diff_unc}
\end{figure*}

We now turn to the discussion of uncertainty estimates for missing higher-order contributions.
In Fig.~\ref{fig:diff_unc} we show the ratio $\rd\sigma_{\rm NNLO}^{\rm QCD}/\rd\sigma_{\rm NLO}^{\rm QCD}$ for the two differential cross sections together with the respective theory-uncertainty estimate for missing QCD corrections coming from standard seven-point scale variations about the central value (see Eq.~\ref{eq:muF_muR}), visualised by the area between the two dashed lines in the respective colours.
Additionally, we provide uncertainty estimates based on the TNP approach introduced in \citere{Lim:2024nsk}, whose extent is shown by the coloured area.
The TNP uncertainties correspond to the 65\% confidence interval.

Focusing first on the transverse-momentum spectrum, we can conclude that the MHOU estimates from scale variations and the TNP approach are of similar magnitude.
For the LT, TL and TT polarisation states, we find MHOU of about $15$-$20$\% at NLO QCD; for the LL state, they are much smaller, of about $2$-$5$\%.
The estimated uncertainties are reduced by a factor of 2-to-3 when going to NNLO QCD, depending on the state.
The TNP approach leads to mildly more conservative estimates and to a slightly improved apparent convergence of the perturbative series, in the sense that the NNLO QCD are more consistent with the NLO QCD uncertainty estimate.
Overall, we find good perturbative convergence for the transverse momentum spectrum.
The $|\Delta y_{\Pe^+\PZ}|$ differential distribution, as discussed above, shows substantial perturbative corrections, particularly in the tail of the distribution.
The theory uncertainty also reflects this, but both estimates underpredict the NNLO QCD corrections in this phase-space regime for the LL, LT and TL polarisation states.
This can be understood again in terms of the effective suppression at tree level, which essentially reduces the perturbative order by 1 in this phase-space region.

We have demonstrated that both NNLO QCD and NLO EW corrections, including PDF-suppressed photon-induced contributions, are sizeable in both inclusive and differential observables, making it important to incorporate them in the SM predictions used in LHC analyses.

\subsection{Joint polarisation fractions}\label{sec:Fresults}

We now switch our focus from polarised cross sections to {\it joint polarisation fractions} which are defined as ratios of doubly polarised cross sections to the unpolarised one (all of them treated in the DPA).
This approach can lead to the sum of fractions that does not equal one, owing to possibly non-vanishing interference effects.
On the contrary, the non-resonant effects (included only in the full off-shell calculation) are considered as an irreducible background to the unpolarised signal.
Therefore, they do not affect the polarisation fractions.

In Table~\ref{tab:fidNNLOcorrelated} we show at various perturbative orders the joint polarisation fractions for the process at hand. While a marked change in the fractions is found between LO ad NLO QCD accuracy, the effect of including NNLO QCD and NLO EW effects is more moderate, with the $\rL\rL$ fraction being nearly unchanged, and a $2\%$ of the $\rT\rT$ contribution being acquired by the mixed states.
The quoted uncertainties in the table are derived from correlated scale variation, i.e. simultaneous variation of scales in the numerator and denominator, and uncorrelated scale variation, i.e. independent variation of the scales in the numerator and denominator. 
\begin{table*}
  \begin{center}
    \begin{tabular}{cccccccc}
      \hline\rule{0ex}{2.7ex}
      \cellcolor{yellow!9} state  
      & \cellcolor{yellow!9} $f_{\rm LO}$ [\%] 
      & \cellcolor{yellow!9} $f^{\rm (QCD)}_{\rm NLO}$ [\%]
      & \cellcolor{yellow!9} $f^{\rm (QCD)}_{\rm NNLO}$ [\%]
      & \cellcolor{yellow!9} $f^{(+)}_{\rm NLO}$ [\%] 
      & \cellcolor{yellow!9} $f^{(\times)}_{\rm NLO}$ [\%]
      & \cellcolor{yellow!9} $f^{(+)}_{\rm NNLO}$ [\%] 
      & \cellcolor{yellow!9} $f^{(\times)}_{\rm NNLO}$ [\%] \\[0.1cm]
      \hline\\[-0.25cm]
off-sh. & $ 2.12  $   & $ 1.88  $   & $ 1.97 $   & $ 1.90  $   & $ 1.89  $   & $ 1.99  $   & $ 1.98 $ \\[0.1cm]
\hline\\[-0.25cm]
unpol. & $ 100 $ & $ 100 $ & $ 100 $ & $ 100 $ & $ 100 $ & $ 100 $ & $ 100 $  \\[0.25cm]
LL & $ 7.88 ^{+0.03 ( + 0.53  )  }_{-0.03 (  -0.66  )  } $   & $ 5.71 ^{+0.17 ( + 0.34  )  }_{-0.18 (  -0.27  )  } $   & $ 5.69 ^{+0.04 ( + 0.18  )  }_{-0.03 (  -0.15  )  } $   & $ 5.66 ^{+0.17 ( + 0.34  )  }_{-0.18 (  -0.28  )  } $   & $ 5.66 ^{+0.17 ( + 0.35  )  }_{-0.18 (  -0.28  )  } $   & $ 5.63 ^{+0.04 ( + 0.19  )  }_{-0.02 (  -0.16  )  } $   & $ 5.64 ^{+0.04 ( + 0.18  )  }_{-0.03 (  -0.18  )  } $ \\[0.25cm]
LT & $ 10.65 ^{+0.10 ( + 0.77  )  }_{-0.12 (  -0.95  )  } $   & $ 15.42 ^{+0.30 ( + 1.38  )  }_{-0.29 (  -1.11  )  } $   & $ 16.46 ^{+0.27 ( + 0.77  )  }_{-0.22 (  -0.65  )  } $   & $ 15.59 ^{+0.30 ( + 1.42  )  }_{-0.29 (  -1.15  )  } $   & $ 15.56 ^{+0.29 ( + 1.43  )  }_{-0.30 (  -1.15  )  } $    & $ 16.65 ^{+0.26 ( + 0.81  )  }_{-0.22 (  -0.68  )  } $   & $ 16.60 ^{+0.28 ( + 0.77  )  }_{-0.24 (  -0.71  )  } $  \\[0.25cm]
TL & $ 10.04 ^{+0.09 ( + 0.72  )  }_{-0.11 (  -0.89  )  } $   & $ 14.68 ^{+0.29 ( + 1.32  )  }_{-0.29 (  -1.06  )  } $   & $ 15.71 ^{+0.22 ( + 0.71  )  }_{-0.20 (  -0.61  )  } $   & $ 15.01 ^{+0.28 ( + 1.37  )  }_{-0.28 (  -1.10  )  } $   & $ 14.98 ^{+0.28 ( + 1.38  )  }_{-0.28 (  -1.11  )  } $   & $ 16.03 ^{+0.21 ( + 0.75  )  }_{-0.19 (  -0.63  )  } $   & $ 15.99 ^{+0.22 ( + 0.70  )  }_{-0.21 (  -0.66  )  } $ \\[0.25cm]
TT & $ 70.67 ^{+0.25 ( + 4.44  )  }_{-0.21 (  -5.58  )  } $   & $ 63.55 ^{+0.40 ( + 4.38  )  }_{-0.45 (  -3.53  )  } $   & $ 61.52 ^{+0.45 ( + 1.72  )  }_{-0.58 (  -1.56  )  } $   & $ 63.11 ^{+0.39 ( + 4.48  )  }_{-0.44 (  -3.60  )  } $   & $ 63.13 ^{+0.39 ( + 4.55  )  }_{-0.43 (  -3.66  )  } $    & $ 61.07 ^{+0.45 ( + 1.84  )  }_{-0.57 (  -1.64  )  } $   & $ 61.11 ^{+0.47 ( + 1.67  )  }_{-0.60 (  -1.81  )  } $ \\[0.25cm]
interf. & $ 0.76  $   & $ 0.63  $   & $ 0.63  $   & $ 0.63  $   & $ 0.67  $   & $ 0.62  $   & $ 0.66  $ \\[0.1cm]
    \hline
    \end{tabular}\qquad
  \end{center}
  \caption{ Fiducial polarisation fractions in the ATLAS setup \cite{ATLAS:2022oge} described in Eq.~\ref{eq:fid}. For the combination of QCD and EW corrections, the same notation as in Table~\ref{tab:predIntegNNLO} is understood. The fraction for each state is defined as the ratio of the corresponding cross-section over the unpolarised one. For polarisation fractions, the absolute QCD-scale uncertainties appear as superscripts and subscripts, and are obtained by varying the central scale in the numerator and in the denominator in a correlated (uncorrelated) manner.
  \label{tab:fidNNLOcorrelated}
  }
\end{table*}

\begin{figure*}
    \centering
    \subfigure[\label{fig:ptzTNPf}]{\includegraphics[width=0.48\textwidth,page=2]{plots/wz_TNP_fract.pdf}}
    \subfigure[\label{fig:dyezTNPf}]{\includegraphics[width=0.48\textwidth,page=1]{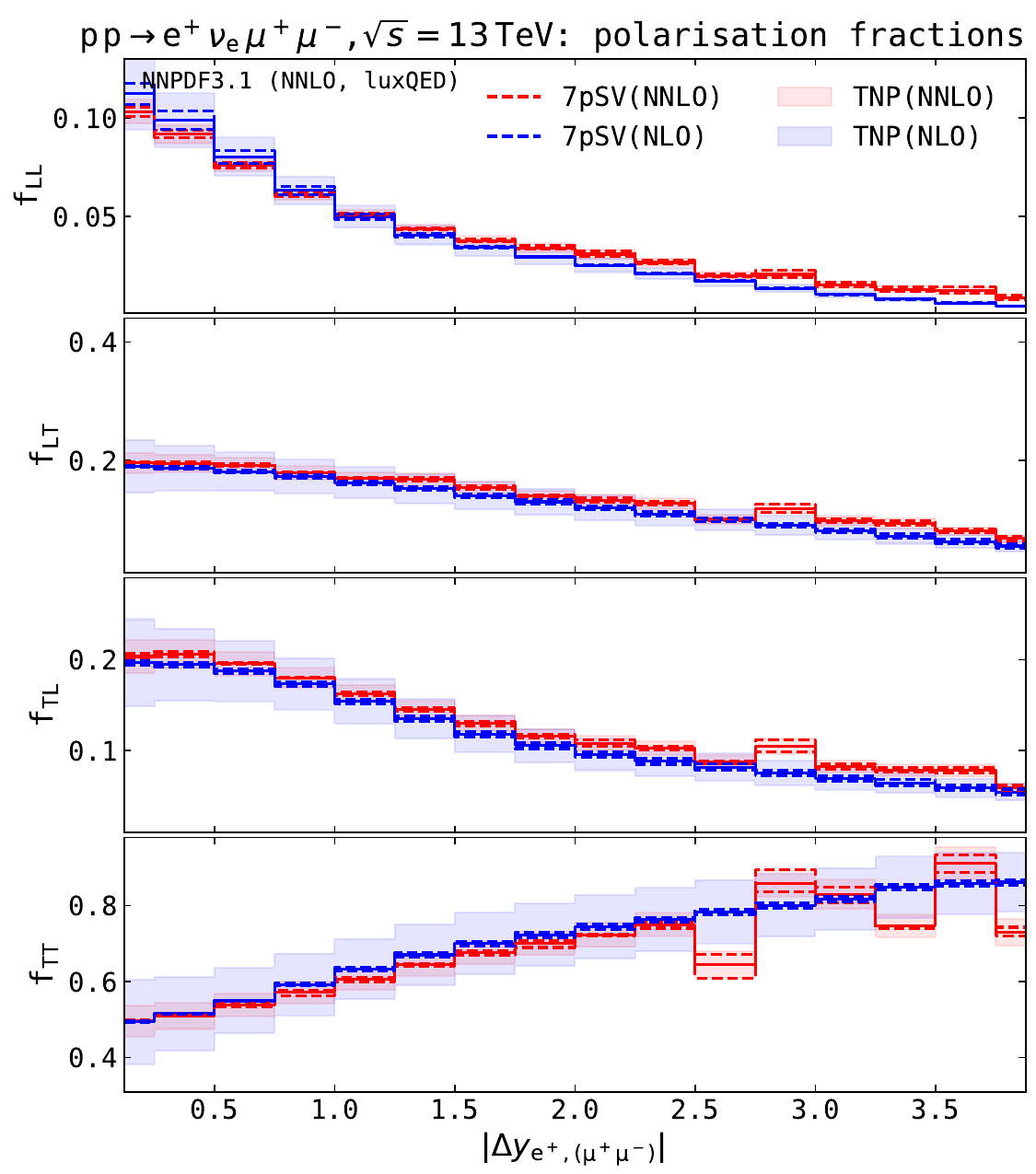}}
    \caption{
    Differential polarisation fractions $f_{\lambda\lambda'}$ ($=\rd\sigma_{\lambda\lambda'}/\rd\sigma_{\rm unp}$) in the transverse-momentum of the muon--antimuon system (a)
    and the rapidity separation between the positron and the muon--antimuon system (b) in the fiducial ATLAS setup \cite{ATLAS:2022oge}. 
    The NLO (NNLO) QCD fractions are shown in blue (red) colour for the LL, LT, TL, and TT states, from top down. The MHOU QCD uncertainties are computed with correlated seven-point scale variations (``7pSV'', dashed boundary lines) and with the theory nuisance parameter approach (``TNP'', shaded bands).
    }
    \label{fig:diff_frac}
\end{figure*}

The differential polarisation fractions for the two observables discussed in the previous section are shown in Fig.~\ref{fig:diff_frac}.

As in the previous section, we show uncertainties estimated from scale variations as the area between the dashed lines, and TNP estimates of MHOU are indicated by the coloured bands.
Scale variations can lead to pathological behaviour in the case of observables defined as ratios of cross sections, due to correlations that do not adequately represent the effects of higher-order corrections.
When considering TNPs to parameterise higher-order corrections, we assume that the corrections to the numerator and the denominator are largely uncorrelated.
This is motivated, for example, by the fact that we can observe that the higher-order corrections differ for doubly polarised and unpolarised configurations.
Thus, we exploit Gaussian error propagation to get an uncertainty $\Delta f_{\lambda\lambda'}^{\rm TNP}$ on the fraction $f_{\lambda\lambda'}$:
\begin{align}
f_{\lambda\lambda'} = \frac{\rm{d} \sigma_{\lambda\lambda'}}{\rm{d}\sigma_{\rm unp}}\,,\quad \text{and}\quad 
\Delta f_{\lambda\lambda'}^{\rm TNP} =  \frac{\rm{d}\sigma_{\lambda\lambda'}}{\rm{d}\sigma_{\rm unp}} \sqrt{
\left(\frac{\rm{d}\Delta \sigma_{\lambda\lambda'}^{\rm TNP}}{\rm{d}\sigma_{\lambda\lambda'}}\right)^2+
\left(\frac{\rm{d}\Delta \sigma_{\rm unp}^{\rm TNP}}{\rm{d}\sigma_{\rm unp}}\right)^2}\;.
\end{align}
One can make several observations in the two differential distributions shown in Fig.~\ref{fig:diff_frac}.
The first observation is that the uncertainties from scale variations essentially vanish for all polarisations, except for the LL state.
Particularly in the tails of the distributions, this does not capture the actual NNLO QCD corrections in the LT, TL, and LL cases.
That the TT polarisation fraction is less sensitive to QCD radiation is expected since the TT polarisation dominates the unpolarised cross section and therefore QCD corrections largely cancel out in the ratio.
The TNP uncertainties are substantially more conservative and comparable to those in the uncorrelated scale-variation case, which we do not show here for clarity.
In this case, uncertainties are significantly reduced when moving from NLO to NNLO QCD.

Finally, we can also compare the TNP uncertainties associated to the fiducial integrated fractions to those obtained from seven-point scale variation. After integrating over the fiducial phase space, the NLO QCD joint fractions with the corresponding TNP uncertainties are found to be,
\begin{align}
f_{\rL\rL} = 0.0571\pm 0.0056\,,\quad
f_{\rL\rT} = 0.1542\pm 0.0236 \,,\quad
f_{\rT\rL} = 0.1468\pm 0.0226\,,\quad
f_{\rT\rT} = 0.6355\pm 0.0724 \,,
\end{align}
while at NNLO QCD they read, 
\begin{align}
f_{\rL\rL} = 0.0569\pm 0.0025\,,\quad
f_{\rL\rT} = 0.1646\pm 0.0113\,,\quad
f_{\rT\rL} = 0.1571\pm 0.0108\,,\quad
f_{\rT\rT} = 0.6152\pm 0.0308\,,
\end{align}
which highlight a more conservative uncertainty estimation compared to the uncorrelated scale variations whose results are detailed in Tab.~\ref{tab:fidNNLOcorrelated}.

\section{Conclusions}\label{sec:conclusion}
We have presented the first calculation of NNLO QCD corrections to doubly polarised $\PW\PZ$ pairs produced inclusively at the LHC and undergoing leptonic decays, and we have combined them with NLO EW effects, achieving the highest perturbative accuracy for this process (same as for the full off-shell description).

The integrated and differential results are shown for a realistic fiducial volume inspired by recent ATLAS analyses. 
The NNLO QCD effects turn out to be sizeable, both at integrated and at differential level, with different impacts on the various doubly polarised signals.
The effect of NLO EW corrections is smaller, but non-negligible,  and becomes of the same size as NNLO QCD corrections in suppressed regions of the phase space.
It is therefore of high importance to combine NNLO QCD and NLO EW corrections, either additively or multiplicatively, to have the best Standard-Model predictions for data interpretation.
Besides the explicitly discussed observables, we provide more numerical results for a broad range of observables as ancillary files.

Our work sets the new state-of-the-art for what concerns fixed-order predictions for polarised $\PW\PZ$ production, but also provides essential building blocks for the matching of NNLO QCD and NLO EW corrections to QCD and QED parton showers.
This is left for future investigations.

Finally, we have investigated a parameterisation of missing higher orders through a theory-nuisance-parameter approach, both for absolute differential polarised predictions and polarised fractions, and compared the results to scale variations.
The found TNP uncertainties tend to be more conservative than scale variations, and also lead to an improvement in the apparent perturbative convergence.
For the polarisation fractions, the TNP strategy is found to be more conservative but also more likely to correctly predict higher-order corrections.

\section*{Acknowledgements}
We would like to thank Ansgar Denner, Lucia Di Ciaccio, Christoph Haitz, and Matthew Lim for useful discussions.
The authors acknowledge support from the COMETA EU COST Action (CA22130).
GP acknowledges financial support from the EU Horizon Europe research and innovation programme under the Marie-Sk\l{}odowska Curie Action (MSCA) ``POEBLITA - POlarised Electroweak Bosons at the LHC with Improved Theoretical Accuracy'' - grant agreement Nr.~101149251 (CUP H45E2300129000) 
and from the Italian Ministry of University and Research (MUR), with EU funds (NextGenerationEU), through the PRIN2022 grant agreement Nr.~20229KEFAM (CUP H53D23000980006). 
RP acknowledges that this research was funded in part by NCN 2024/55/D/ST2/00934.
This work was performed in part using the Cambridge Service for Data Driven Discovery (CSD3), part of which is operated by the University of Cambridge Research Computing on behalf of the STFC DiRAC HPC Facility (www.dirac.ac.uk). The DiRAC component of CSD3 was supported by STFC grants ST/P002307/1, ST/R002452/1 and ST/R00689X/1.

\appendix
\section{Comparison with the literature}\label{app:compNinh}
We provide the numerical comparison of joint polarisation fractions with \citere{Le:2022lrp}.
The ATLAS setup \cite{ATLAS:2022oge} described in Eq.~\ref{eq:fid} is understood.
Slightly different input SM parameters and different PDF sets are used for the two setups.
\begin{table*}[htb]
  \begin{center}
    \begin{tabular}{ccccc}
      \hline\rule{0ex}{2.7ex}
       \cellcolor{yellow!9} & \multicolumn{2}{c}{\cellcolor{yellow!9} $f^{\rm (QCD)}_{\rm NLO}$ [\%]} & \multicolumn{2}{c}{\cellcolor{yellow!9} $f^{(+)}_{\rm NLO}$ [\%]} \\[0.1cm]
      \hline\\[-0.2cm]
       \cellcolor{yellow!9}  
       & \cellcolor{yellow!9}this work
      & \cellcolor{yellow!9}\cite{Le:2022lrp}
      & \cellcolor{yellow!9}this work
      & \cellcolor{yellow!9}\cite{Le:2022lrp}\\[0.1cm]
      \hline\\[-0.25cm]
LL   
& $ 5.71 $ & 5.7
& $ 5.66 $ &  5.6\\[0.25cm]
LT   & $ 15.42  $& 15.5
& $ 15.59  $ &  15.6 \\[0.25cm]
TL  & $ 14.68  $ &  14.7   & $ 15.01  $ &  15.1 \\[0.25cm]
TT   & $ 63.55  $&  63.5  & $ 63.11  $  & 63.0 \\[0.15cm]
    \hline
    \end{tabular}\qquad
  \end{center}
  \caption{ Comparison of fiducial polarisation fractions with \citere{Le:2022lrp}. 
  \label{tab:fNLOcomp}
  }
\end{table*}

\bibliographystyle{JHEPmod}
\bibliography{polvv}

\end{document}